\documentstyle[twoside,fleqn,espcrc2]{article}

\title{Instanton size distributions from calibrated cooling}

\author{C.~Michael\address{DAMTP, University of Liverpool, PO Box 147,
        Liverpool L69 3BX, UK} 
	and 
	P.S.~Spencer\address{Research Institute for Theoretical
        Physics (TFT), Siltavuorenpenger 20C, 00014 Helsinki,
        Finland.}  \thanks{Talk given by PSS. Work carried 
        out as part of the EC Programme 
        ``Human Capital and Mobility''---project number
        ERB-CHRX-CT92-0051.}}
       
\begin{document}

\begin{abstract}
Using an under-relaxed cooling algorithm we investigated the vacuum in
the $2d\ O(3)$ model and $4d$ pure gauge $SU(2)$. We calibrated the
amount of cooling performed to have similar physical effect at
different lattice spacings.

[Liverpool preprint: LTH 337; TFT preprint: HU-TFT-94-45, hep-lat/9411015]
\end{abstract}

\maketitle

\section{Introduction}
Topology is an important aspect of many lattice field theories, in
particular QCD, where a non-zero value of the topological
susceptibility is necessary to solve the $U(1)$ problem and explain
the mass of the $\eta'$ meson.  For QCD the topology is
a consequence of the existence of instanton solutions of the theory,
so simpler theories than QCD that share the property of instanton
solutions have long been of interest.

In this paper we present work carried out on two such simpler theories:
$2d~O(3)$ and $4d$ pure gauge $SU(2)$.  In both of these theories we
extracted information about the underlying long-range structure of the
theory by cooling, a process which has been widely used for this
purpose.  It was immediately apparent to us that, as we wished to
discuss the underlying physics, we would need to compare results
obtained at different lattice spacings and we would therefore need to
calibrate the cooling we used to have the same {\em physical\/} effect
across the range of $a$ that we used; performing the same, arbitrary
number of cooling sweeps at each value of $a$ would have different
physical effects at different couplings, rendering a comparison of the
results meaningless---different cooling levels would mean the gross
structure of the configuration would be different, and it is on this
structure that we wish to make calculations and compare results. 

\section{The $O(3)$ Model}
The $2d\ O(3)$ model shares many characteristics with QCD: it is
asymptotically free, with dynamic mass generation and instanton
solutions due in this case to the homotopy class of windings from
$S_2\rightarrow S_2$. 

Using a standard stereographic projection the single instanton
solution is given by:
\begin{equation}\label{e:1o3i}
\omega(z,\overline{z})=\frac{\phi_1+{\rm
i}\phi_2}{1-\phi_3}=\frac{\rho}{z-r}
\;,\;z=x+{\rm i}t
\end{equation}
\noindent with both $\rho$ and $r$ complex, and $\phi_i$ the $i$th
component of $\phi$. The action of this
continuum field configuration is $S_I=4\pi/g^2$. (It should be noted
that on a torus there is no single-instanton solution~\cite{usdist}.)
In general multi-instanton solutions are given by analytic functions
of $\omega$, ie $\partial_{\overline{z}}\omega=0$.

Unfortunately, the lattice version of the $O(3)$ model has problems,
as the theory is dominated by short-range fluctuations~\cite{luscale},
and formulating the theory on a lattice imposes a minimum size to
these fluctuations, and so the important contribution from those
smaller than $a$ is lost. Furthermore, additional unphysical
contributions arise from lattice artefacts of size ${\cal O}(a)$. So
on the one hand results obtained from cooling would not have the
contribution from the physical ultra-violet fluctuations, but on the
other hand they would not be corrupted by the unphysical artefacts,
and measurements of objects of moderate size, $a\ll\rho\ll L$, should
be reliable. On these grounds we decided to investigate the size
distribution of instantons.

We used an under-relaxed cooling given by:
\begin{equation}\label{e:o3cool}
\phi_x'=\alpha\phi_x + \phi^F_x\;,\;
\mbox{$\phi^F_x=\sum_{\mu}(\phi_{x+\mu}+\phi_{x-\mu})$}
\end{equation}
\noindent with $\phi_x'$ normalised to unit length. This was used for
two reasons: firstly, as a deterministic update it is much faster to
compute, and secondly: the parameter $\alpha$ allows us to control the
severity of the cooling, though a trade-off must be made between the
gentler cooling associated with larger values of $\alpha$ and the
greater number of sweeps then needed to reach a given physical
state. We eventually decided that $\alpha=2$ was a good compromise
between too many sweeps and too harsh a cool. ${\cal O}(a)$ effects in
the lattice $O(3)$ action mean that lattice instantons are unstable
under cooling and will be shrunk and eventually annihilated by
prolonged cooling. We decided to calibrate our cooling for the $O(3)$
model by measuring the number of cooling sweeps at different values of
$\alpha$ required to annihilate a configuration generated from a
discretised form of eq.~\ref{e:1o3i}. We looked at three levels of
cooling at each of three couplings: $g^2=1.00, 0.84, 0.80$
corresponding to mass gaps of $ma=0.551(1), 0.381(1), 0.261(1)$, the
levels being those required to remove objects of size one-third, one
half and two-thirds the correlation length $\xi=1/m$ at each value of
$g^2$. The number of sweeps needed are given in
table~\ref{t:o3cool}. We found that $S/S_IV$ calculated on the cooled
configurations was consistent across $g^2$ for each of the cooling
levels we used, and took this as evidence for correct calibration of
our cooling.

\begin{table}[t]
\setlength{\tabcolsep}{0.91pc}
\newlength{\digitwidth} \settowidth{\digitwidth}{\rm 0}
\catcode`?=\active \def?{\kern\digitwidth}
\caption{Details of the cooling used for $O(3)$. Data calculated on a
$64^2$ lattice.}
\label{t:o3cool}
\begin{center}
\begin{tabular}{ccccc}
\hline
$g^2$ & $ma$ &$\xi/3$&$\xi/2$&$2\xi/3$\\ \hline
0.80 & 0.261(1) & 28 & 79 & 220\\
0.84 & 0.318(1) & 17 & 45 & 106\\
1.00 & 0.551(1) &  5 & 13  & 24 \\ 
\hline
\end{tabular}
\end{center}
\end{table}

\section{Pure gauge $SU(2)$}
For $SU(2)$ we again used an under-relaxed cooling, this time given
by:
\begin{equation}\label{e:su2cool}
U_{x,\mu}'=\alpha U_{x,\mu}+\Sigma_{x,\mu}
\end{equation}
\noindent with $U_{x,\mu}'$ normalised to lie in $SU(2)$ and
$\Sigma_{x,\mu}$ the sum of the `staples' around $U_{x,\mu}$.

When we came to extend the techniques we developed for $O(3)$ to the
case of $4d\ SU(2)$ we ran into the problem that the lattice $SU(2)$
instantons are much more stable under cooling than their $O(3)$
counterparts---so stable in fact that had we na\"\i vely taken the
same criterion as earlier, we would have cooled the configurations
beyond any region where we would wish to study the vacuum, and in
certain cases removed {\em all\/} the physics present in the uncooled
configuration. In light of this we changed our approach: as we wish to
consider the vacuum as quantum fluctuations around classical
solutions, it makes sense, taking into account the stability of the
lattice instantons, to calibrate the cooling by first constructing
classical solutions, introducing quantum fluctuations by a few Monte
Carlo updates and then cooling until the original configuration is
recovered. This procedure is performed a number of times and the
average number of sweeps needed to recover the initial state is taken
as the calibration. Preliminary studies of this on $16^4$ lattices at
$\beta=2.4$ and $24^4$ at $\beta=2.5$ have proved promising, and the
data imply that 61 sweeps at $\alpha=2$ is sufficient to remove the
quantum fluctuations without overly disturbing the underlying
structure on a $16^4$ lattice at $\beta=2.4$.

\section{Instanton Size Distributions}
\begin{figure}[hb]
\vspace{5 in}
   \includegraphics{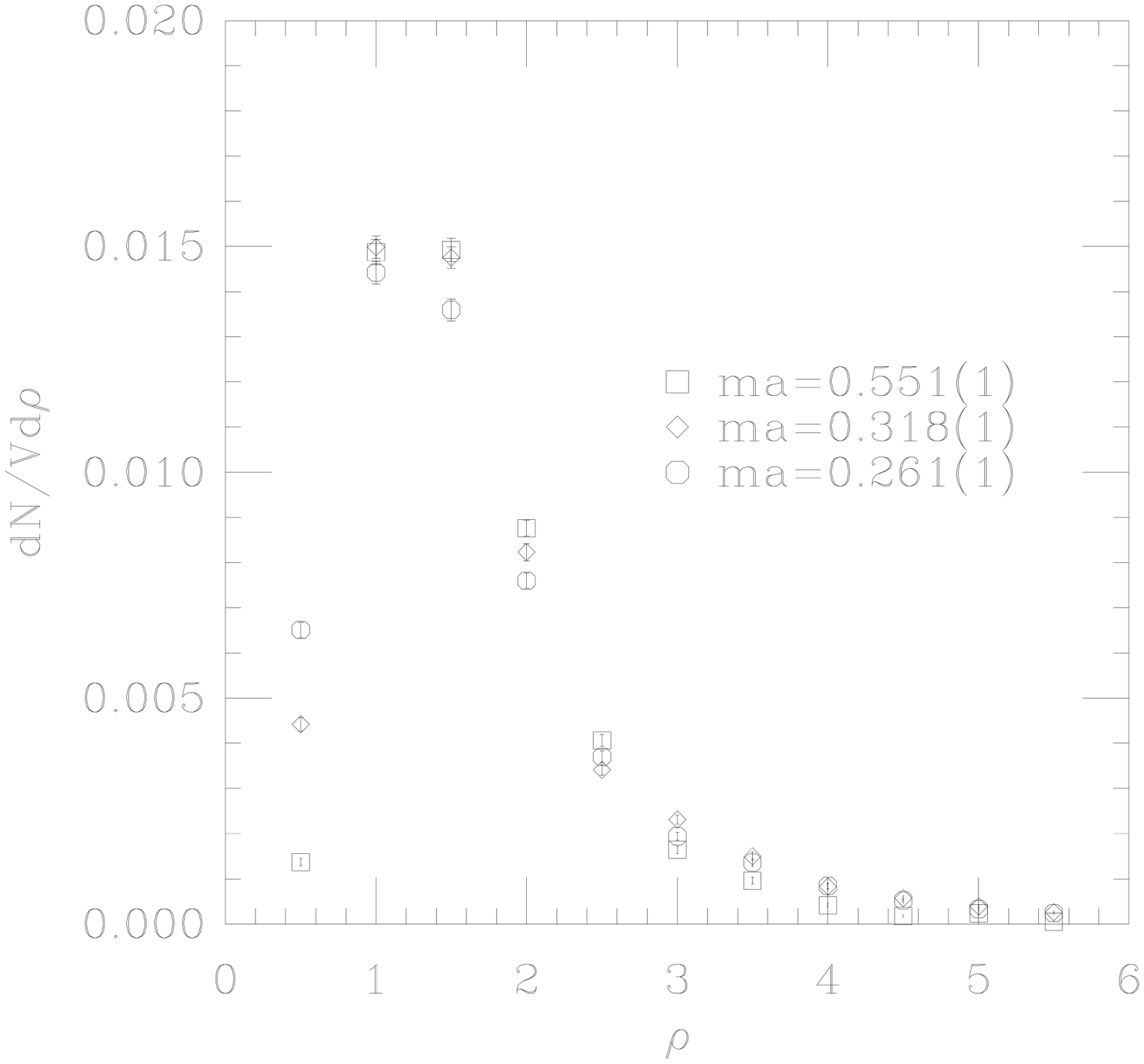}
   \includegraphics{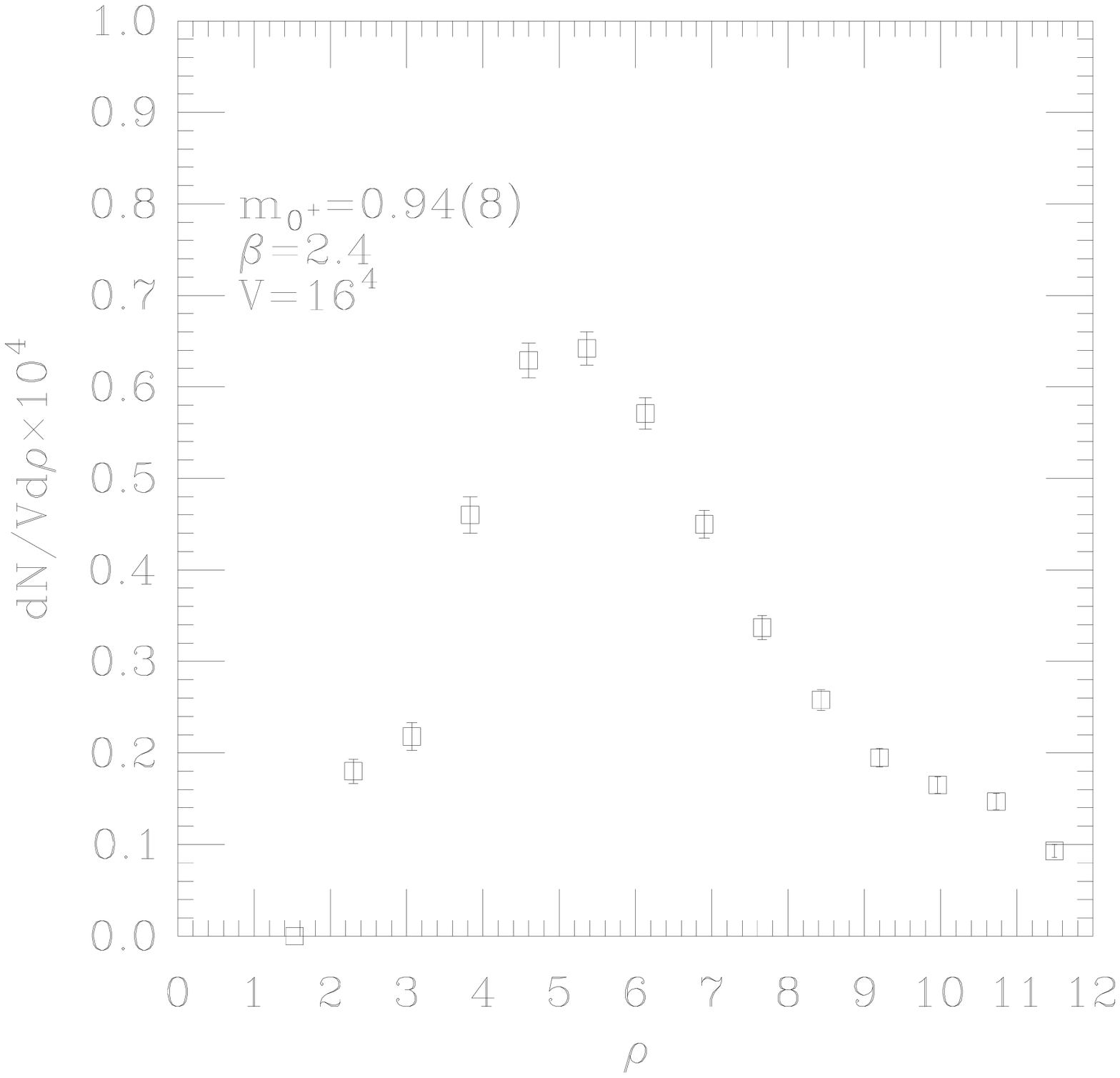}
\vspace{1 in}
\caption{\label{f:dists}
The distributions of instanton size calculated for $O(3)$ from 1000
configurations for each value of $a$ on a $64^2$ lattice (upper plot),
and for $SU(2)$ from 1000 configurations at $\beta=2.4$ on a $16^4$
lattice (lower plot). Instanton sizes are given in units of the
correlation length, $\xi=1/m$; for $SU(2)$,
\mbox{$m=m_{0^+}$.} We define $N=S/S_I$}
\end{figure}

In order to calculate the sizes of instantons we looked for connected
regions around local maxima in the action density for which the action
density was not less than half the value at the maximum under
consideration; 2-dimensional regions in $O(3)$ and 4-dimensional for
$SU(2)$. We then took the appropriate root of this volume to obtain a
size parameter. Predictions of the instanton size distribution in both
$O(3)$ and $SU(2)$ can be derived from the dilute instanton gas model
(see, for example, ref.~\cite{rajaraman}). For $O(3)$ the model
predicts
\begin{equation}
\frac{1}{S_IV}\frac{dS}{d\rho}\sim\frac{1}{\rho}
\end{equation}
\noindent whereas our data, presented in~\cite{usdist} and summarised
in figure~\ref{f:dists}, indicate a much stronger UV dependence and a
distribution $\sim\rho^{-3}$. For $SU(2)$, the dilute gas predicts an
infra-red divergence and 
\begin{equation}
\frac{1}{S_IV}\frac{dS}{d\rho}\sim\rho^{7/3}
\end{equation}
\noindent It is unclear whether our preliminary data, shown in
figure~\ref{f:dists} are in agreement with this. Certainly they
indicate no ultra-violet divergence, and it appears that the very
smallest instantons are all but absent.

For $O(3)$, while the finite lattice spacing reduces any signal below
$\rho\sim 1.5\xi$, we see a rapid decrease at larger $\rho$.  For $SU(2)$
this decrease is less rapid. The physical volumes here are
approximately equal, with $L \sim 16\xi$, so we take this difference
in large $\rho$ behaviour to be a physical effect.

For $O(3)$ we also looked at the separation of pairs of instantons and
anti-instantons, and found evidence for an interaction: the closest
separation of unlike pairs of objects was significantly smaller than
that of like pairs. At each value of $g^2$ we found that unlike pairs
had a closest separation only 70\% that of like pairs. This
interaction, and the form of the distribution we found show no support
for the dilute gas model for $O(3)$ instantons on the lattice. The
corresponding $SU(2)$ measurements are in progress.

\section{Conclusions}
We have given an argument in favour of calibrating cooling to have a
given physical effect, and altering the number of sweeps as the
lattice spacing is altered.  We have also presented data calculated
using calibrated cooling and shown how it is consistent across a range
of couplings.

\end{document}